\documentclass[12pt]{article}
\usepackage{amsmath}
\usepackage{float}
\usepackage{xcolor}
\usepackage{indentfirst}
\usepackage{setspace}
\usepackage{graphicx}

\usepackage[left=2.5cm,right=2.5cm,top=2.5cm,bottom=2.5cm]{geometry}
\begin{document}


\begin{center}
{\Large
	{\textbf{Choosing good subsamples for regression modelling}}
}
\vskip1.5em

 \textbf{Thomas Lumley  and Tong Chen}\footnote{Thomas Lumley, Department of Statistics, University of
  Auckland, Auckland, New Zealand, 1010. E-mail: t.lumley@auckland.ac.nz; Tong Chen, Department of Statistics, University of
  Auckland, Auckland, New Zealand, 1010. E-mail: tche929@aucklanduni.ac.nz. } 

\end{center}

\begin{center}
    {\bf Abstract}
\end{center}

\noindent A common problem in health research is that we have a large database with many variables measured on a large number of individuals. We are interested in measuring additional variables on a subsample; these measurements may be newly available, or expensive, or simply not considered when the data were first collected. The intended use for the new measurements is to fit a regression model generalisable to the whole cohort (and to its source population). This is a two-phase sampling problem; it differs from some other two-phase sampling problems in the richness of the phase I data and in the goal of regression modelling. In particular, an important special case is measurement-error models, where a variable strongly correlated with the phase II measurements is available at phase I.  We will explain how influence functions have been useful as a unifying concept for extending classical results to this setting, and describe the steps from designing for a simple weighted estimator at known parameter values through adaptive multiwave designs and the use of prior information.  We will conclude with some comments on the information gap between design-based and model-based estimators in this setting.


\onehalfspacing

\section{Introduction}

Two-phase sampling, where a phase I sample is taken and then a subsample taken using the phase-I data, has a history dating back at least to Neyman (1938).  In health research there are increasingly many examples where the phase-I data are rich, and there are relatively few variables being measured at phase II, and examples where the goal of data collection is to fit regression models.  Settings for this sort of two-phase sampling include cohort studies, where new or expensive variables are measured on a subset of cohort members, and outcomes research, where the phase-II data come from labour-intensive audits of a subsample of routinely collected health records (Baldoni et al., 2021). Case--control studies, a core tool of epidemiology, can be seen as an intermediate step, lacking rich phase-I data but being concerned primarily with regression modelling. 

In this paper we discuss optimal design of subsamples, primarily for design-based analysis and in the context of health research. Most subsampling of existing cohorts in health research has not involved a focus on efficient design beyond the use of case--control and case--cohort sampling. There are a few examples, such as the two NIH/NHLBI-funded sequencing `HeartGO' initiatives (Lin et al., 2014; Tao et al., 2015).

Choosing the optimal subsample is also desirable in validating  routinely collected health records.
Electronic health record (EHR) data are not collected for research purposes. They are prone to errors: actual mistakes and mismatches between administrative and research definitions of variables. Human chart reviews can correct some of the error-prone data, but are expensive and time-consuming. Performing the validation on a well-selected subsample gives the benefits of chart review at lower cost.  Recently, Shepherd et al. (2021) studied the association of maternal weight gain during pregnancy with child risk factors using the EHR data.  A subsample of records, which are selected by the optimal design for design-based analysis, is validated by manual chart review.

After considering notation, we describe the use of influence functions in section~\ref{influence}, and then consider the dependence of the optimal design on unknown parameters. This dependence is approached by multiwave sampling (section~\ref{multiwave}), and by the use of priors (section~\ref{prior}).

Most current work on optimal design-based inference optimises the design for a simple inverse-probability-weighted estimator.  In practice, a generalised raking (or AIPW) estimator will be used; section~\ref{raking} discusses this gap. Model-based analysis is an important alternative approach, with analysis either by semiparametric maximum likelihood (Tao et al., 2020a; Lee et al., 2001; Scott and Wild, 2001) or by multiple imputation (eg Giganti and Shepherd, 2020). In section~\ref{model} we discuss the differences between optimal designs for design-based and model-based analysis.  Finally, in section~\ref{software} we comment on software availability.

\subsection{Notation}
We are working in the intersection of multiple fields of statistics, which complicates notation. For example, in the measurement error literature it is common for $W$ to refer to a version of a covariate $X$ that is measured with error; in the survey world, $W$ would be weights.  

Here is our suggestion, which appears to be reasonably compatible with other fields.  We sample 
$n$ observations from a cohort or database of $N$, where the $i$th observation is sampled with known probability $\pi_i$  and sampling indicator $R_i$. The sampling weights $w_i$ are $1/\pi_n$ or adjusted versions of this to incorporate cohort-level information.
We have variables $Z$, $A$, and (typically) $Y$ measured for everyone in the cohort and $X$ measured on the subsample.  The \emph{outcome model} is for $Y|Z,X$. It is the model we would fit if we had complete data. Its parameters are $\beta$; its log-likelihood for a single observation is $\ell_i(\beta)$; its score function for a single observation is $U_i(\beta)$. 

The \emph{imputation model} is for $X|Z,A,Y$. Its parameters are $\alpha$. It may be used to produce single imputations $\hat X_i$ or multiple imputations $\hat X_i^{(m)}$ for $m=1,2,\dots, M$. The \emph{phase-1 model} is for $Y|\hat X, Z$. It has influence functions $h_i(\beta)$ (or for multiple imputation, $h_i^{(m)}(\beta)$). We use the term raking (or generalised raking) for the process of adjusting weights using full-cohort information, to avoid confusion with the unrelated `regression calibration' technique in the measurement-error literature.

\section{Influence functions: from sums to parameters}
\label{influence}

A unifying concept in translating classical design and analysis results for sums to regression modelling is the \emph{influence function}. In the full cohort, if we estimate a parameter $\theta$ by the census parameter $\tilde\theta$, we can write
$$\sqrt{n}(\tilde\theta-\theta )= \frac{1}{\sqrt{N}}\sum_{i=1}^n h_i(\theta)+o_p(1)\label{inf-fun-unwt},$$
where $h_i(\theta)$ is a function only of observation $i$.  For a weighted estimator $\hat\theta$ based on a subsample, we can write
\begin{align}
    \sqrt{n}(\hat\theta-\tilde\theta )= \frac{1}{\sqrt{n}}\sum_{i=1}^n R_iw_ih_i(\theta)+o_p(1).\label{inf-fun-wt}
\end{align}

If $\tilde\theta$ solves unbiased estimating equations
$$\sum_{i=1}^N U_i(\theta)=0,$$
then (under some regularity conditions)
$$h_i(\theta) = -\left(\sum_i \frac{\partial U_i}{\partial \theta}\right)^{-1} U_i(\theta).$$
More generally,  Demnati and Rao (2004) showed the influence functions can be computed as the derivatives of $\hat\theta$ with respect to weights on each observation. 

The use of estimating functions to estimate standard errors for parameter estimates is well known in survey statistics (Binder, 1983;  Rao et al., 2002) and in biostatistical modelling.  We and co-workers have previously found that thinking in terms of influence functions illuminates the choice of auxiliary variable for raking when the target of inference is a regression parameter (Lumley et al., 2011; Breslow et al., 2009), though the same estimators can be developed in other ways (Robins et al., 1994). In this paper we argue that influence functions also provide a useful heuristic in optimal sampling design. 

Under stratified random sampling for estimation of the population total of some variable $Y$, Neyman showed the optimal allocation of a fixed number of observations across $K$ strata (Neyman, 1934) was proportional to the with-stratum standard deviation of $Y$ and to the population size in the stratum.  Neyman allocation typically does not give integer values, but Wright (2012) provided an exact integer algorithm.  These allocation rules can be applied to estimation of a regression parameter $\beta$, because the regression parameter estimate is asymptotically equivalent to the estimated population total of its influence functions as shown in Equation~(\ref{inf-fun-wt}).  That is, the optimal allocation of observations at phase II to strata is proportion to the size of the strata and the standard deviation of $h_i(\beta)$ (Chen and Lumley, 2020). In the case of a binary $X$, $Y$, and $A$, McIsaac and Cook (2015) arrived at this design rule by direct optimisation using Lagrange multipliers; they did not comment on the relationship to Neyman allocation.

\subsection{Example: case--control design}

The case--control design provides a simple illustration of the use of influence functions in design.  Suppose that at phase I we have just a binary $Y$, and that we measure $X$ at phase II. We wish to fit a logistic regression model 
$$\mathrm{logit} P[Y=1|X=x]=\mathrm{logit} p_x=\beta_0+x\beta_x.$$
The prevalence of $Y=1$ is low. Not only is $p_0=E[Y]$ small, $p_x$ is small for all observable $x$; this is the setting where the case--control design is useful. We will take a sample stratified on $Y$. 

The influence function in the model is proportional to the score function $U_i=x_i(y_i-p_i)$. In the $Y=1$ stratum
$$\mathrm{var}[U_i]=\mathrm{var}[X(Y-p_x)]\approx\mathrm{var}[X|Y=1]\approx\mathrm{var}[X],$$
where the last approximate equality would be exact if $\beta=0$. In the control stratum
$$\mathrm{var}[U_i]=\mathrm{var}[X(0-p_x)]\approx p_0^2\mathrm{var}[X].$$

The Neyman allocation rule says to take $n$ for a stratum proportional to $N\sqrt{\mathrm{var}[U_i]}$, which is approximately $Np_0\sqrt{\mathrm{var}[X]}$ for cases and $N(1-p_0)\sqrt{p_0^2\mathrm{var}[X]}$ for controls, which are approximately equal. So, the influence-function approach gives a simple heuristic for the 1:1 case--control design, as well as allowing detailed optimisation of the design to the extent that more detailed information is available. 

\section{Unknown parameters}

In general, the optimal design will depend on both the parameters $\beta$ in the outcome model and on the parameters $\alpha$ in the imputation model, which are unknown, and so cannot be evaluated in practice. The optimal design still serves as a benchmark for evaluating achievable designs. In this section we consider two approaches to approximating the optimal design: multiwave sampling and the use of Bayesian priors.

\subsection{Multiwave sampling}
\label{multiwave}
McIsaac and Cook (2015) proposed multiwave sampling to approximate the optimal design.  In this approach, a small subsample is taken in wave I and used to estimate the unknown parameters $\alpha$ and $\beta$.  The design is then optimised for the estimated values $\hat\alpha$ and $\hat\beta$ and the remainder of the sample is taken in wave II. Clearly this design could also be generalised to more than two waves, allowing for information to be gathered incrementally throughout the sampling process. 

When wave I is too small, the estimates of $\alpha$ and $\beta$ will not be accurate enough. When wave I is large enough, the estimates of $\alpha$ and $\beta$ will be accurate enough to work out an approximately optimal design. However, if wave I is too large, the number sampled in some strata in wave I may already be larger than the number required by the optimal design, leading to sub-optimality, so that appropriate choice of wave I size is important.

Shepherd et al. (2021) studied
maternal and child risk factors for childhood obesity and asthma in 10,335 children, using electronic health record data augmented by a Phase II sample of 996 manually audited records. A six-wave sampling design was used to choose a validation sample for gestational age, weight gain during pregnancy, and asthma status. Approximately half of records (500 records) were validated at the first two waves to ensure that the estimates of $\alpha$ and $\beta$ were accurate. The other half of records were validated adaptively through four waves.

\subsection{Priors}
\label{prior}
Even though $\alpha$ and $\beta$ will not be known accurately before data are collected, it is also unlikely that there will be no prior information about their values. In the context of McIsaac and Cook (2015), the unknown parameters measure the association between an exposure and outcome, and the accuracy of a biomarker for that exposure.  When research has progressed to the extent that an expensive two-phase study makes sense, domain experts are likely to know something about the plausible ranges of $\alpha$ and $\beta$.

Chen and Lumley (2020) investigated the extent to which formal prior distributions over $\alpha$ and $\beta$ can guide the design of a multiwave study in the setting given by McIsaac and Cook. We did not carry out a formal Bayesian {\it analysis}; the analysis was entirely design based. However, we found that moderately informative priors (variance of 1 on the log odds scale) will reduce the probability of a poor design and will allow smaller wave I samples to be taken, even when the prior is not centred unrealistically close to the true parameter.  The use of a formal prior  appears to be helpful both by providing extra information and by regularising the estimates from the first wave. 

We generated 1000 phase-I samples with size $N = 1000$ and sampled $n=300$ at phase II. Suppose $X$ followed a Bernoulli distribution with $\mathrm{Bern}(0.15)$. A surrogate variable $A$ was generated with sensitivity $0.8$ and specificity $0.8$. We also generated a continuous variable $Z_1 \sim \mathrm{U}(0,1)$ and a binary variable $Z_2 \sim \mathrm{Bern}(0.6)$. The outcome variable $Y$ was generated with 
\begin{align*}
     \mathrm{logit}(p_y) &= -2 + 0.5 X + Z_1 + Z_2,\\
    Y &\sim \mathrm{Bern} (p_y),
\end{align*}
where $\mathrm{logit}(p_y) = \log (p_y/(1-p_y))$. The data were divided into $8$ strata based on $A$, $Y$, and $Z_2$. We compared the following two-wave designs 
\begin{enumerate}
    \item A two-wave sampling with well-calibrated strong priors (well.strong) where $\beta_i \sim N(\beta_i-\sqrt{0.1}/2, 0.1)$, $\alpha_j \sim N(\alpha_j-\sqrt{0.1}/2, 0.1)$;
    \item A two-wave sampling with well-calibrated weak priors (well.weak) where $\beta_i \sim N(\beta_i-\sqrt{0.1}/2, 1)$, $\alpha_j \sim N(\alpha_j-\sqrt{0.1}/2, 1)$; 
    \item A two-wave sampling with poorly-calibrated strong priors (poor.strong) where $\beta_i \sim N(\beta_i-1/2, 0.1)$, $\alpha_j \sim N(\alpha_j-1/2, 0.1)$; 
    
    \item A two-wave sampling with poorly-calibrated weak priors (poor.weak) where $\beta_i \sim N(\beta_i-1/2, 1)$, $\alpha_j \sim N(\alpha_j-1/2, 1)$;
    
    \item A two-wave sampling with proportional stratified sampling at wave I (prop.two).
\end{enumerate}

\begin{figure}[H]
\centering
\scalebox{0.8}{\includegraphics[width=0.9\textwidth]{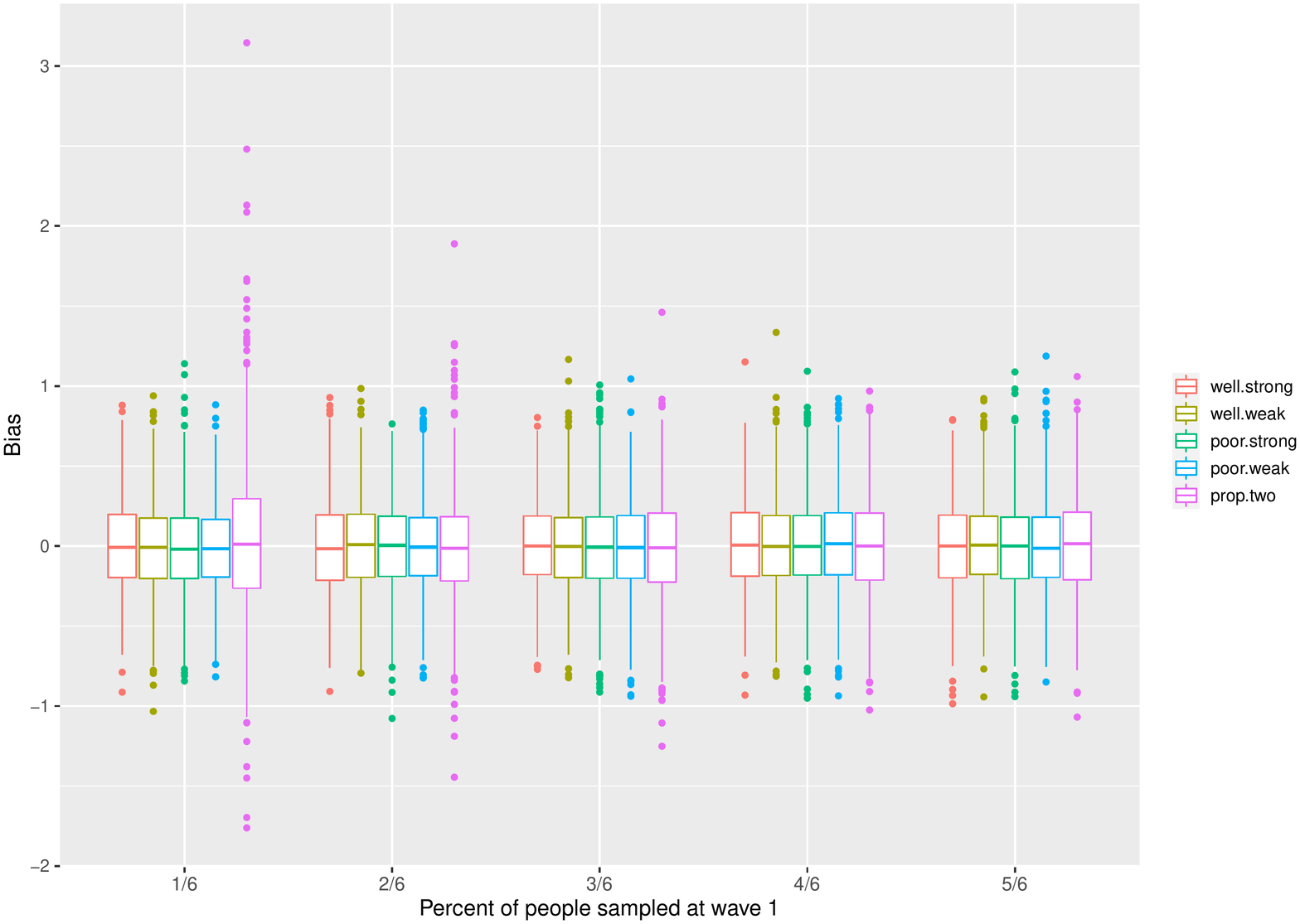}}
\caption{\footnotesize Bias estimated from generalised raking estimators for different two-wave designs.}\label{f1}
\end{figure}

\begin{figure}[H]
\centering
\scalebox{0.8}{\includegraphics[width=0.9\textwidth]{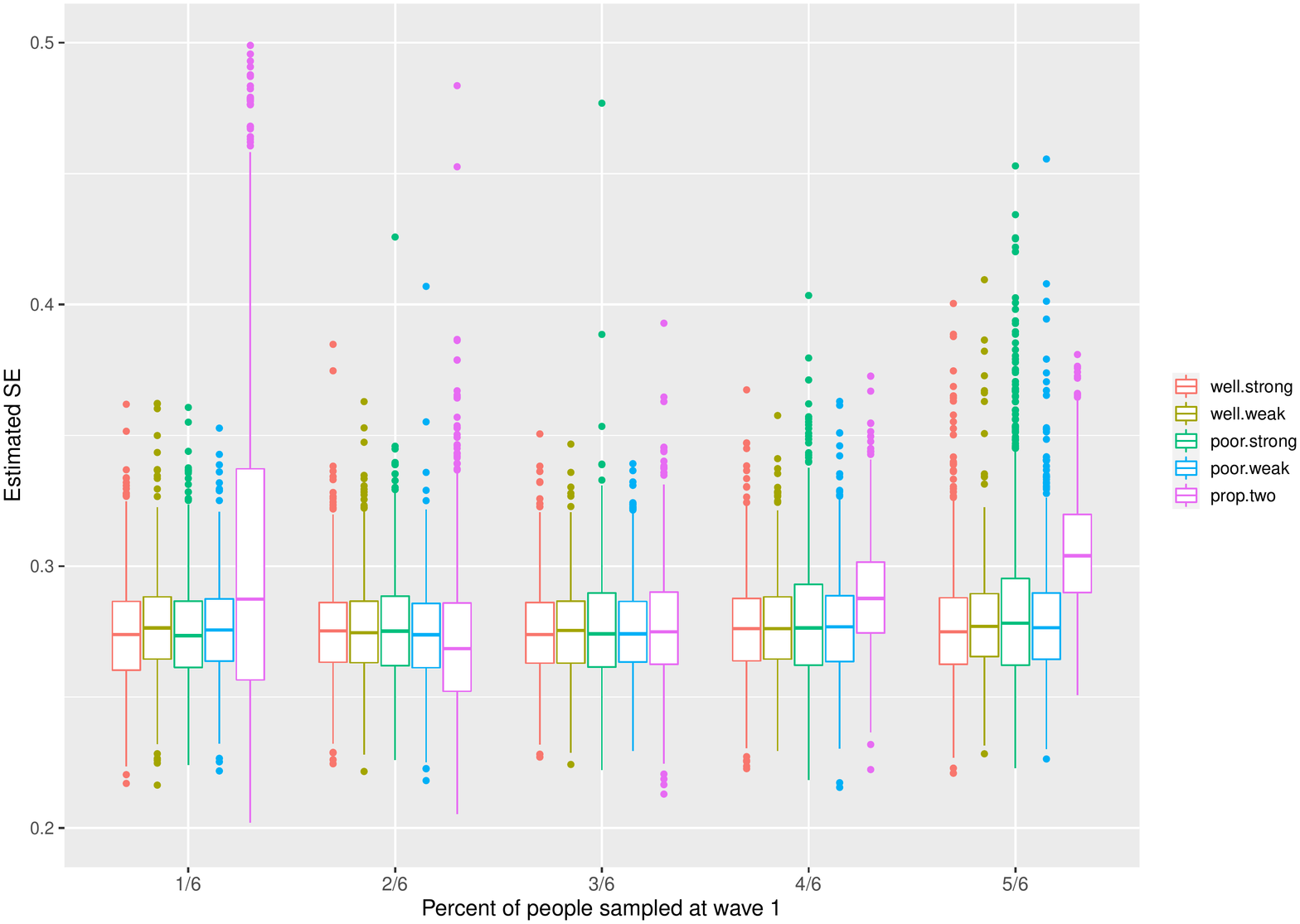}}
\caption{\footnotesize Estimated standard error of $\hat{\beta}$ estimated from generalised raking estimators for different two-wave designs.}\label{f2}
\end{figure}

Figure~\ref{f1} and Figure~\ref{f2} show that two-wave designs with priors over $\alpha$ and $\beta$ will be more efficient than that with proportional stratified sampling at wave I, especially when wave-I sample size is small. Chen and Lumley (2020) found broadly similar results in a range of scenarios.

\subsection{Optimisation for raking}
\label{raking}

All the previous design optimisation was targeted at simple inverse-probability weighted estimation. In practice, we will use generalised raking to adjust the weights; it will often be possible at analysis time to get some useful prediction of the phase-II influence functions from phase-I data.  This raises a question: how much worse are our designs than the optimal designs for raking? 

When auxiliary variables are good, without model assumptions, there is not much room for the efficiency gain from design optimisation. When auxiliary variables are bad, since the correlation between auxiliary variables and influence functions is low, adjusting the weights based on raking will not improve on the inverse-probability weighted estimation. In practice, auxiliary variables are less likely to be either good or bad; it is still possible to improve from optimising the design for generalised raking.

The optimal allocation for a regression parameter after raking follows the same principles as for an IPW estimator, the only change is that the stratum-specific standard deviation of the influence functions is replaced by a stratum-specific standard deviation conditional on the raking variables (Chen and Lumley, 2022). More precisely, if $h_i(\beta)$ are the influence functions and $a_i$ are the auxiliary variables, the relevant standard deviation is of the residuals $r_i$ obtained from linear regression of $h_i(\beta)$ on $a_i$.

Unfortunately, it is difficult to estimate the standard deviation of $r_i$ at design time.  The auxiliary variables $a_i$ should already be our best estimates of $h_i$ from phase-I data, and until the data have been collected we do not have accurate values $h_i$ to regress on them. All we can do is compare the optimal designs for raking and IPW estimators under parametric assumptions, in order to learn how the designs differ in these specific settings.  It turns out that the optimal design for raking is typically closer to proportional allocation than the optimal design without raking for linear regression, but that the two designs typically lead to similar variances for the regression parameters (Chen and Lumley, 2022). For logistic regression, the optimal design for raking is very close to the optimal design without raking. That is, optimising the design for IPW estimation is often sufficient; as it is much easier, this is very fortunate (Chen and Lumley, 2022). 

We generated 2000 phase-I samples with size $4000$ and sampled $n = 600$ at phase II. Suppose both $X$ and $Y$ were continuous where $X$ followed a standard normal distribution. An error-prone variable $\tilde{X}$ was generated with $\tilde{X} = X+U$, where $\mathrm{U} \sim N(0, 0.5^2)$. The outcome variable $Y$ was generated with $Y = 1 + 0\times X + Z_1 +Z_2+\epsilon$ where $Z_1 \sim \mathrm{Bern}(0.5)$ and $Z_2$ and $\epsilon$ both followed a standard normal distribution. We defined 3 strata based on the cut-off points at the 20th and 80th percentiles of $\tilde{X}$. We compared the optimal design for analysis via the IPW estimator (IF-IPW) with that for analysis via the generalised raking estimators (IF-GR) under generalised raking estimations. 

\begin{figure}[H]
\centering
\scalebox{0.75}{\includegraphics[width=0.9\textwidth]{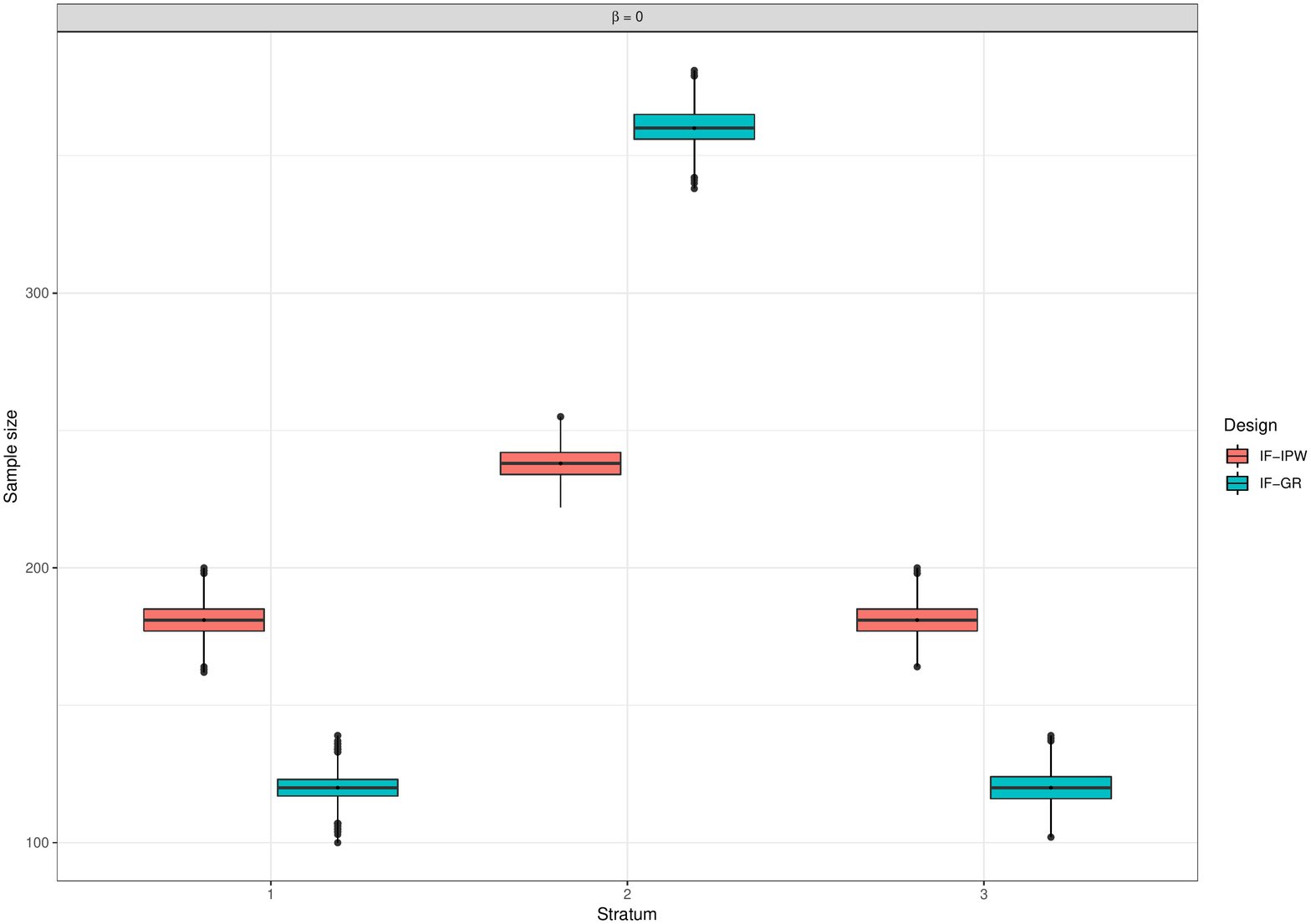}}
\caption{\footnotesize Empirical comparison of sample sizes between the optimal design for analysis via the IPW estimator (IF-IPW) and those for analysis via generalized raking estimators (IF-GR).}\label{f3}
\end{figure}

\begin{figure}[H]
\centering
\scalebox{0.75}{\includegraphics[width=0.9\textwidth]{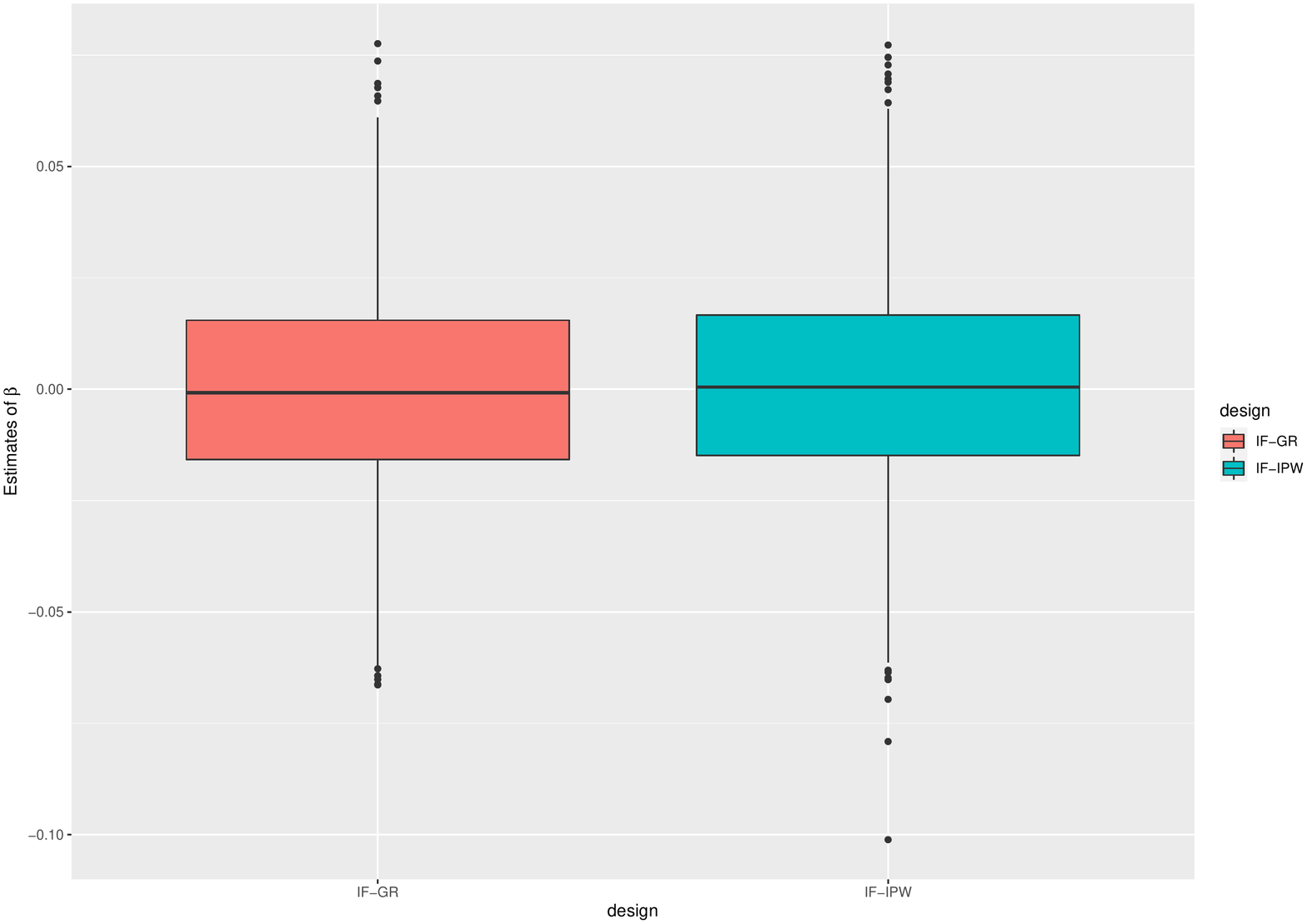}}
\caption{\footnotesize Empirical comparison of parameter $\beta$ estimated from raking analysis between IF-IPW and IF-GR. }\label{f4}
\end{figure}

Figure~\ref{f3} and Figure~\ref{f4} show that the two designs end up with very similar efficiency under generalised raking analysis, though the two designs are different. Broadly the same results were found for a wide range of scenarios in Chen and Lumley (2022).

\section{Model-based analysis: the efficiency gap}
\label{model}

Model-based analysis can in principle be much more efficient than model-assisted design-based analysis.  Design-based analysis can be viewed mathematically as analysis under the model that assumes correctly-specified sampling probabilities, and nothing else. This model is much larger than those assumed by model-based analysis, and the semiparametric information bound can consequently be much higher.  On the other hand, the information gap need not always be large, and even when it is large there is a genuine precision/robustness tradeoff in making stronger assumptions.

A convenient illustration that the information gap may be large, but need not be,  comes from case--control sampling.  With a single binary covariate, design-based inference is fully efficient; with a single Normal covariate the efficiency ranges from 100\% at an odds ratio of 1 down to well below 50\% for large odds ratios.  Lumley and Scott (2017) analyse a classic case--control data set and show that design-based inference is nearly efficient in the widely-used discrete version of the data, but quite inefficient in the original continuous version.

The local asymptotic minimax theorem (eg van der Vaart, 1997, Chapter 8) implies that model-based analyses cannot be superior to design-based analyses \textit{uniformly over neighbourhoods of contiguous alternatives}; there must exist some hard-to-detect model misspecification under which the model-based estimator has higher asymptotic mean squared error than the best design-based estimator.  This result does not imply that the worst-case misspecification will be plausible, but the analyses of Han et al. (2021) provide examples where it is, and Breslow et al. (2013) give an illustration in data from the Women's Health Initiative.  Amorim et al. (2021) compared several model-based and design-based strategies and found that they all gave improvements over simple random sampling, and argued that the ideal choice of design and analysis depended on context.

In some special cases, optimal sampling strategies for model-based and design-based estimators are the same. Nedyalkova and Tillé (2008) studied the optimal design and estimation for predictions among linear estimators, and showed that under balanced sampling with sampling probabilities proportional to the standard deviations of the errors of the model, if population model has fully explainable heteroscedasticity, the best linear unbiased predictor equals the Horvitz-Thompson estimator, so the optimal designs are the same. The optimal designs of model-based and design-based estimators are also the same for logistic regression under case--control sampling with a rare outcome at $\beta = 0$.

The optimal designs based on model-based estimators and design-based estimators can be very different. As is familiar in experimental design, the optimal sampling of continuous predictors often involves taking extreme values of the predictors and not sampling intermediate values.  Design-based inference is not possible under these designs; many of the sampling probabilities are zero.  Tao et al. (2020b) developed locally optimal designs for maximum-likelihood analysis when the regression coefficients are close to zero and claimed that these designs also work well for moderate non-zero associations; these designs typically have zero sampling probabilities for some units in the population.

Optimal designs for model-based estimators with model-based estimation can in principle be much more efficient than those for design-based estimators under correct model specification, but this may not be true if the model is misspecified. The minimax design has been studied in some literature. Clark (2020) showed the lower bound for the anticipated variance of model-assisted estimators is the asymptotic upper bound for that of the best linear unbiased predictor under a linear model and probability sampling, which implies the optimal design for analysis via model-assisted estimators is minimax for model-based estimators. Welsh and Wiens (2013) studied the minimax design for model-based estimators and stated that points are spread throughout the design space under the minimax design. Future research is merit to explore when the design optimality is sensitive to mild model misspecification.

\section{Software}
\label{software}
Statistical methods need software. The {\sf survey} package for R (Lumley, 2004; Lumley, 2020) provides survey analysis facilities, including analysis of two-phase samples with raking/calibration of weights, and the {\sf sampling} package (Till\'e and Matei, 2021) carries out sampling, including PPS and covariate-balanced designs. Neither package provides much specific support for design of two-phase samples from existing cohorts. 

Published software for optimising two-phase designs included those of Schill and Wild (2006), who constructed minimax designs and Haneuse et al. (2011), who used simulation to evaluate specific choices of design. The ``2-Phase Planning Tool" of Schill et al. (2007) was designed to simplify the choice of stratification and evaluation of alternatives; it did not perform any optimisation. 

Yang and Shaw (2021) have developed an R package to simplify optimal two-phase and adaptive multi-wave design. It performs optimal allocation to strata using Neyman allocation or Wright's exact allocation. More importantly, the package allows for straightforward merging and splitting of strata to improve a design.

\section{Conclusions}

There is increasing interest in subsampling from existing cohorts and databases in health research, and in doing so efficiently.  While theoretical development is ahead of current practice in many ways, it is still not clear how best to sample when the optimal design depends on unknown parameters. Optimal designs can be very different for  model-assisted and model-based estimation, and the practical implications of these differences also require more research and deeper understanding.

\section*{Acknowledgments}
This work was supported in part by Patient Centered Outcomes Research Institute
(PCORI) Award R-1609-36207, U.S. National Institutes of Health (NIH) grant R01-
AI131771, and the University of Auckland doctoral scholarship (to the second author). The statements in this manuscript are solely the responsibility of the authors and do not necessarily represent the views of PCORI or NIH.

\bigskip
\bigskip
\begin{center}
{\bf \Large References}
\end{center}
\parskip16pt
\noindent Amorim, G., Tao, R., Lotspeich, S., Shaw, P.A., Lumley, T. and Shepherd, B.E. (2021). Two-phase sampling designs for data validation in settings with covariate measurement error and continuous outcome. \textit{Journal of the Royal Statistical Society, Series $A$}, 184, 1368--1389. 

\noindent Baldoni, P.L., Sotres-Alvarez, D., Lumley, T. and Shaw, P.A. (2021). On the use of regression calibration in a complex sampling design with application to the Hispanic Community Health Study/Study of Latinos. \textit{American Journal of Epidemiology}, 190, 1366–-1376.

\noindent Binder, D.A. (1983). On the variance of asymptotically normal estimators from complex surveys. \textit{International Statistical Review / Revue Internationale de Statistique}, 51, 279--292.

\noindent Breslow, N.E., Amorim, G., Pettinger, M.B. and  Rossouw, J. (2013). Using the whole cohort in the analysis of case-control data: Application to the Women's Health Initiative. \textit{Statistics in Biosciences}, 5, 232--249.

\noindent Breslow, N.E., Lumley, T., Ballantyne, C.M., Chambless, L.E. and  Kulich, M. (2009). Using the whole cohort in the analysis of case-cohort data. \textit{American Journal of Epidemiology}, 169, 1398--1405. 

\noindent Chen, T., and Lumley, T. (2020).  Optimal multiwave sampling for regression modeling in two-phase designs. \textit{Statistics in Medicine}, 39, 4912--4921.

\noindent Chen, T., and Lumley, T. (2022).  Optimal sampling for design-based estimators of regression models. \textit{Statistics in Medicine}, 41, 1482--1497.

\noindent Clark, R.G. (2020). Model-assisted sample design is minimax for model-based prediction.  \textit{Survey Methodology,} 46, 77–-91.

\noindent Demnati, A., and Rao, J.N.K. (2004). Linearization variance estimators for survey data. \textit{Survey Methodology}, 30, 17--26.

\noindent Giganti M.J., and Shepherd B.E. (2020). Multiple-imputation variance estimation in studies with missing or misclassified inclusion criteria. \textit{American Journal of Epidemiology}, 189, 1628--1632.

\noindent Han, K., Shaw, P.A. and  Lumley T. (2021).  Combining multiple imputation with raking of weights: An efficient and robust approach in the setting of nearly true models. \textit{Statistics in Medicine}, 40, 6777--6791.
 
\noindent  Haneuse, S., Saegusa, T. and  Lumley T. (2011). osDesign: An R package for the analysis, evaluation, and design of two-phase and case-control studies. \textit{Journal of Statistical Software}, 43, 1--29.

\noindent Lee, A.J., Scott, A.J. and Wild, C.J. (2001). Efficient estimation in multi-phase case-control studies. \textit{Biometrika}, 97, 361--374.

\noindent  Lin, H.,  Wang, M.,  Brody, J.A. and Bis, J.C. and others. (2014). Strategies to design and analyze targeted sequencing data: cohorts for Heart and Aging Research in Genomic Epidemiology (CHARGE) Consortium Targeted Sequencing Study. \textit{Circulation: Cardiovascular Genetics}, 7, 335--343.

\noindent Lumley, T. (2004). Analysis of complex survey samples. \textit{Journal of Statistical Software}, 9, 1--19.

\noindent Lumley, T. (2020). \textit{survey: analysis of complex survey samples}. R package version 4.0. 

\noindent Lumley, T., and  Scott, A. (2017). Fitting regression models to survey data. \textit{Statistical Science}, 32, 265--278.

\noindent Lumley, T., Shaw, P.A. and Dai, J.Y. (2011). Connections between survey calibration estimators and semiparametric models for incomplete data. \textit{International Statistical Review / Revue Internationale de Statistique}, 79, 200--232.

\noindent McIsaac, M.A., and  Cook, R.J. (2015). Adaptive sampling in two-phase designs: a biomarker study for progression in arthritis. \textit{Statistics in Medicine}, 34, 2899--2912.

\noindent Nedyalkova, D., and Till\'e, Y. (2008). Optimal sampling and estimation strategies under the linear model. \textit{Biometrika}, 95, 521--537.

\noindent  Neyman, J. (1934).  On the two different aspects of the representative method: the method of stratified sampling and the method of purposive selection. \textit{Journal of the Royal Statistical Society}, 97, 558--606.

\noindent Neyman, J. (1938). Contribution to the theory of sampling human populations. \textit{Journal of the American Statistical Association}, 33, 101-–116.


\noindent Rao, J.N.K., Yung, W. and Hidiroglou, M.A. (2002). Estimating equations for the analysis of survey data using poststratification information. \textit{Sankhy\={a}: The Indian Journal of Statistics, Series A}, 64, 364--378.

\noindent Robins J.M., Rotnitzky A. and Zhao L.P. (1994).  Estimation of regression coefficients when some regressors are not always observed. \textit{Journal of the American Statistical Association}, 89, 846--866.

\noindent Schill, W., and Wild, P. (2006).  Minmax designs for planning the second phase in a two-phase case-control study. \textit{Statistics in Medicine}, 25, 1646--1659.

\noindent  Schill, W., Wild, P. and Pigeot, I. (2007). A planning tool for two-phase case-control studies. \textit{Computer Methods and Programs in Biomedicine}, 88,  175--181.

\noindent Scott, A.J., and Wild, C.J. (2001). Maximum likelihood for generalised case-control studies. \textit{Journal of Statistical Planning and Inference}, 96, 3--27.

\noindent Shepherd, B.E., Han, K., Chen, T., Bian, A., Pugh, S., Duda, S.N., Lumley, T., Heerman, W.J. and Shaw, P.A. (2021). Analysis of error-prone electronic health records with multi-wave validation sampling: Association of maternal weight gain during pregnancy with childhood outcomes. \textit{arXiv preprint}, arXiv:2109.14001.

\noindent Tao, R., Lotspeich, S.C., Amorim, G., Shaw, P.A. and Shepherd, B.E. (2020a). Efficient semiparametric inference for two-phase studies with outcome and covariate measurement errors. \textit{Statistics in Medicine}, 40, 725--738.

\noindent Tao, R., Zeng, D., Franceschini, N., North, K.E., Boerwinkle, E. and Lin, D.Y. (2015). Analysis of sequence data under multivariate trait-dependent sampling. \textit{Journal of the American Statistical Association}, 110, 560--572.

\noindent Tao, R., Zeng, D. and Lin, D.Y. (2020b). Optimal designs of two-phase studies. \textit{Journal of the American Statistical Association,} 115, 1946--1959.

\noindent Till\'e, Y., and Matei, A. (2021). \textit{sampling: survey sampling}. R package version 2.9.

\noindent van der Vaart, A.W. (1998). \textit{Asymptotic Statistics}. Cambridge University Press.

\noindent Welsh, A.H., and Wiens, D.P. (2013). Robust model-based sampling designs. \textit{Statistics and Computing,} 23, 689–-701.

\noindent Wright, T. (2012). The equivalence of Neyman optimum allocation for sampling and equal proportions for apportioning the U.S. House of Representatives. \textit{The American Statistician}, 66, 217--224.

\noindent Yang, J., and Shaw P.A. (2021). \textit{optimall: allocate samples among strata}. R package version 0.1. \texttt{https://github/yangjasp/optimall}.

\end{document}